\documentclass[runningheads]{llncs}
\usepackage{graphicx}
\usepackage{cite}
\usepackage{amsmath,amssymb,amsfonts}
\usepackage{textcomp}
\usepackage{xcolor}
\usepackage{booktabs}
\usepackage[hidelinks]{hyperref}
\usepackage{subfig}
\usepackage{wrapfig}
\usepackage{algpseudocode}
\usepackage{algorithm}
\usepackage{multirow}
\usepackage{graphicx}
\def\BibTeX{{\rm B\kern-.05em{\sc i\kern-.025em b}\kern-.08em
    T\kern-.1667em\lower.7ex\hbox{E}\kern-.125emX}}

\algrenewcomment[1]{\(\triangleright\) #1}
\algnewcommand{\LineComment}[1]{\State \(\triangleright\) #1}
\usepackage[utf8]{inputenc}

\begin{document}
\title{FedCC: Robust Federated Learning against Model Poisoning Attacks}
\titlerunning{FedCC}
% \author{\em Anonymous Authors}
\author{Hyejun Jeong \inst{1}, Hamin Son \inst{2} \and
Seohu Lee\inst{3} \and \\
Jayun Hyun \inst{4} \and Tai-Myoung Chung \inst{4}}
\authorrunning{H. Jeong et al.}
\institute{UMass Amherst 
\email{hjeong@umass.edu}
\and{UC Davis 
\email{sonhamin3@gmail.com}
\and{Johns Hopkins University 
\email{slee619@jhu.edu}
\and{Hippo T\&C Inc. 
\email{\{jayunhyun, tmchung\}@skku.edu}}}}}
%

% \author{\IEEEauthorblockN{1\textsuperscript{st} Hyejun Jeong}
% \IEEEauthorblockA{\textit{Computer Science} \\
% \textit{UMass Amherst}\\
% Amherst, USA \\
% hjeong@umass.edu}
% \and
% \IEEEauthorblockN{2\textsuperscript{nd} Ha Min Son}
% \IEEEauthorblockA{\textit{Computer Science} \\
% \textit{UC Davis}\\
% Davis, USA \\
% sonhamin3@gmail.com}
% \and
% \IEEEauthorblockN{3\textsuperscript{rd} Seohu Lee}
% \IEEEauthorblockA{\textit{School of Medicine} \\
% \textit{Johns Hopkins University}\\
% Baltimore, USA \\
% slee619@jhu.edu}
% \and
% \IEEEauthorblockN{4\textsuperscript{th} Jayun Hyun}
% \IEEEauthorblockA{\textit{Hippo T\&C Inc.}\\
% Suwon, South Korea \\
% jayunhyun@g.skku.edu}
% \and
% \IEEEauthorblockN{5\textsuperscript{th} Tai-Myoung Chung}
% \IEEEauthorblockA{\textit{Hippo T\&C Inc.} \\
% Suwon, South Korea \\
% tmchung@skku.edu}
% }

\maketitle              % typeset the header of the contribution
\begin{abstract}
Federated learning is a distributed framework designed to address privacy concerns. However, it introduces new attack surfaces, which are especially prone when data is non-Independently and Identically Distributed. Previous approaches often tackle non-IID data and poisoning attacks separately. To address both challenges simultaneously, we present FedCC, a simple yet effective novel defense algorithm against model poisoning attacks. It leverages the Centered Kernel Alignment similarity of Penultimate Layer Representations for clustering, allowing the identification and filtration of malicious clients, even in non-IID data settings. The penultimate layer representations are meaningful since the later layers are more sensitive to local data distributions, which allows better detection of malicious clients. The sophisticated utilization of layer-wise Centered Kernel Alignment similarity allows attack mitigation while leveraging useful knowledge obtained.
Our extensive experiments demonstrate the effectiveness of FedCC in mitigating both untargeted model poisoning and targeted backdoor attacks. Compared to existing outlier detection-based and first-order statistics-based methods, FedCC consistently reduces attack confidence to zero. Specifically, it significantly minimizes the average degradation of global performance by 65.5\%. We believe that this new perspective on aggregation makes it a valuable contribution to the field of FL model security and privacy. Code is available at https://github.com/HyejunJeong/FedCC. 


\end{abstract}

\section{Introduction}

Federated Learning (FL) \cite{mcmahan2017communication} is a distributed model training framework designed to preserve privacy by restricting data to remain on client devices. Only model parameters are exchanged, minimizing data privacy risks typically associated with centralized learning. This makes FL particularly useful in data-sensitive environments, where raw data should never leave clients, reducing the risk of data leakage.

However, its distributed nature makes FL susceptible to model poisoning attacks \cite{kairouz2021advances}. The server cannot directly examine local datasets' data quality or model parameter integrity, leaving compromised clients or attackers to manipulate local models. This can degrade global model performance indiscriminately (untargeted attacks) \cite{fang2020local, baruch2019little, shejwalkar2021manipulating} or cause incorrect predictions on specific inputs (targeted attacks) \cite{wang2020attack, xi2021batfl, bhagoji2019analyzing, xie2020dba}. Backdoor attacks are stealthier targeted attacks that maintain overall performance while misclassifying specific inputs \cite{bagdasaryan2020backdoor}.

While some defenses rely on robust aggregation methods, many fail to maintain privacy by sharing raw data or exposing data distributions to untrusted parties \cite{cao2020fltrust, wang2022flare}. Furthermore, most existing defenses assume IID (Independent and Identically Distributed) data, which is rare in real-world scenarios. Non-IID data, where clients' data distributions and features vary, complicates attack detection and performance maintenance \cite{kairouz2021advances, zhang2021federated, yoshida2020hybrid}. As the degree of non-IID data increases, the impact of attacks also grows \cite{li2022improved, shejwalkar2021manipulating}.

In this study, we introduce FedCC, a defense mechanism against untargeted model poisoning and targeted backdoor attacks in both IID and non-IID settings. FedCC leverages Centered Kernel Alignment (CKA) of Penultimate Layer Representations (PLRs) to distinguish malicious clients from benign ones. PLRs are highly sensitive to local data distributions, making clients distinguishable, especially in non-IID environments. By exploiting CKA's ability to measure similarity in high-dimensional spaces, we can accurately identify malicious clients. Empirical evidence, shown in Figure \ref{fig:cluster_distance} and Table \ref{tab:diffsim}, demonstrates that PLRs provide the highest separability between benign and malicious clients, with CKA delivering superior performance in both IID and non-IID settings.

Importantly, FedCC ensures data privacy by relying on similarity measures between models instead of shared data. The proposed method outperforms existing defenses based on first-order statistics, which typically suffer from high false negative rates. Our experiments show that FedCC significantly improves global accuracy against untargeted attacks (66.27\% improvement) and reduces attack confidence in targeted backdoor attacks to nearly zero while preserving main task accuracy.

Our contributions are summarized as follows:
\begin{itemize} 
    \item We propose FedCC, a novel and scalable defense method that uses CKA of PLRs and performs layer-wise weighted aggregation of model parameters to defend against model poisoning attacks. 
    \item We demonstrate that PLRs provide a highly distinguishable feature for detecting malicious clients by measuring the discrepancy between clusters of clients. 
    \item We justify the use of CKA as an accurate and sensitive similarity measure for comparing models, even in non-IID settings, when the server has no access to client data. 
    \item We empirically validate the effectiveness of FedCC through extensive experiments, showing that it outperforms existing defenses, especially in non-IID scenarios. 
\end{itemize}

\section{Backgrounds and Related Works} \label{Section2}

\subsection{Poisoning Attacks in FL}
FL is vulnerable to poisoning attacks due to its distributed nature, where the server cannot directly inspect dataset quality or model integrity. Model poisoning attacks are particularly destructive, as adversaries can stealthily manipulate local model parameters to degrade global model performance \cite{bhagoji2019analyzing, bouacida2021vulnerabilities}, and this paper focuses on such attacks.

In model poisoning attacks, client-side adversaries alter local model parameters before submitting them to the server. To avoid detection and prevent global model divergence, attackers optimize the local models for both training loss and an adversarial objective \cite{bouacida2021vulnerabilities, fang2020local}. 
For instance, A3FL \cite{zhang2024a3fl} dynamically fine-tunes backdoor triggers to make them harder to detect, while IBA \cite{nguyen2024iba} generates robust triggers using a generative network, exploiting the global model as a discriminator.
Attackers can further manipulate hyperparameters, such as learning rate, local epochs, batch size, and regularization \cite{li20233dfed, lyu2023poisoning}, dynamically before and during local training to evade detection. 
In a similar vein, DBA \cite{xie2020dba} embedded split triggers into local training data, associating them with targeted incorrect predictions. Unlike centralized backdoor attacks, DBAs distribute malicious updates across multiple adversarial participants, making them harder to detect with anomaly detection techniques.

Model poisoning attacks can be classified into untargeted and targeted types. Untargeted attacks degrade global model accuracy, while targeted attacks mislead the model to misclassify specific inputs without affecting other classes. Our defense mechanism mitigates both untargeted attacks by restoring global model accuracy and targeted backdoor attacks by overcoming the stealthiness and detection difficulty \cite{xi2021batfl, wang2020attack}.

\subsection{Robust Aggregation algorithm in FL}
Several Byzantine-robust aggregation methods, based on summary statistics or anomaly detection, have been proposed to mitigate model poisoning and backdoor attacks in FL. Krum \cite{blanchard2017machine}, for instance, selects the update with the smallest Euclidean distance to others but assumes IID data and overlooks outliers \cite{baruch2019little}. Multi-Krum, which averages updates from multiple clients, is known to be more effective in non-IID settings \cite{blanchard2017machine}. Similarly, Median \cite{yin2018byzantine} computes the coordinate-wise median, Trimmed Mean \cite{yin2018byzantine} excludes extreme values, and Bulyan \cite{guerraoui2018hidden} combines Krum and Trimmed Mean for added robustness. These methods, however, require knowledge of the number of attackers, often unavailable in practice.

Inspired by the observation that malicious clients’ model parameters exhibit higher similarity, Foolsgold \cite{fung2020limitations} identified Sybils by measuring cosine similarity between gradients. While effective against multiple Sybils, it struggled when only a single malicious client exists or with IID data, where it can overfit. FLTrust \cite{cao2020fltrust} addressed this by assigning trust scores based on ReLU-clipped cosine similarities but risking privacy by raw data sharing. Similarly, Lockdown \cite{huang2024lockdown} improved robustness by pruning parameters unused by the majority of clients during training, while FLIP \cite{zhang2022flip} rejected low-confidence samples at test time in addition to the adversarial training.

Despite these advances, many methods rely on summary statistics (mean or median), direction (cosine similarity), or distance (Euclidean) of weight vectors. However, they often struggle to differentiate benign clients with non-IID data from malicious ones, leading to misclassification and suboptimal performance. The limited performance might be partially attributed to the inadequacy of Euclidean geometry for comparing neural network representations.

In contrast, our approach uses Kernel CKA, a kernel-based metric, to measure the similarity between global and local models. This method effectively identifies compromised model parameters among clients with diverse data distributions, enhancing defense against model poisoning in FL and overcoming the shortcomings of previous techniques.

\subsection{Penultimate Layer Representation in FL}

\begin{figure}
    \vspace{-5mm}
    \centering
    \includegraphics[width=0.9\linewidth]{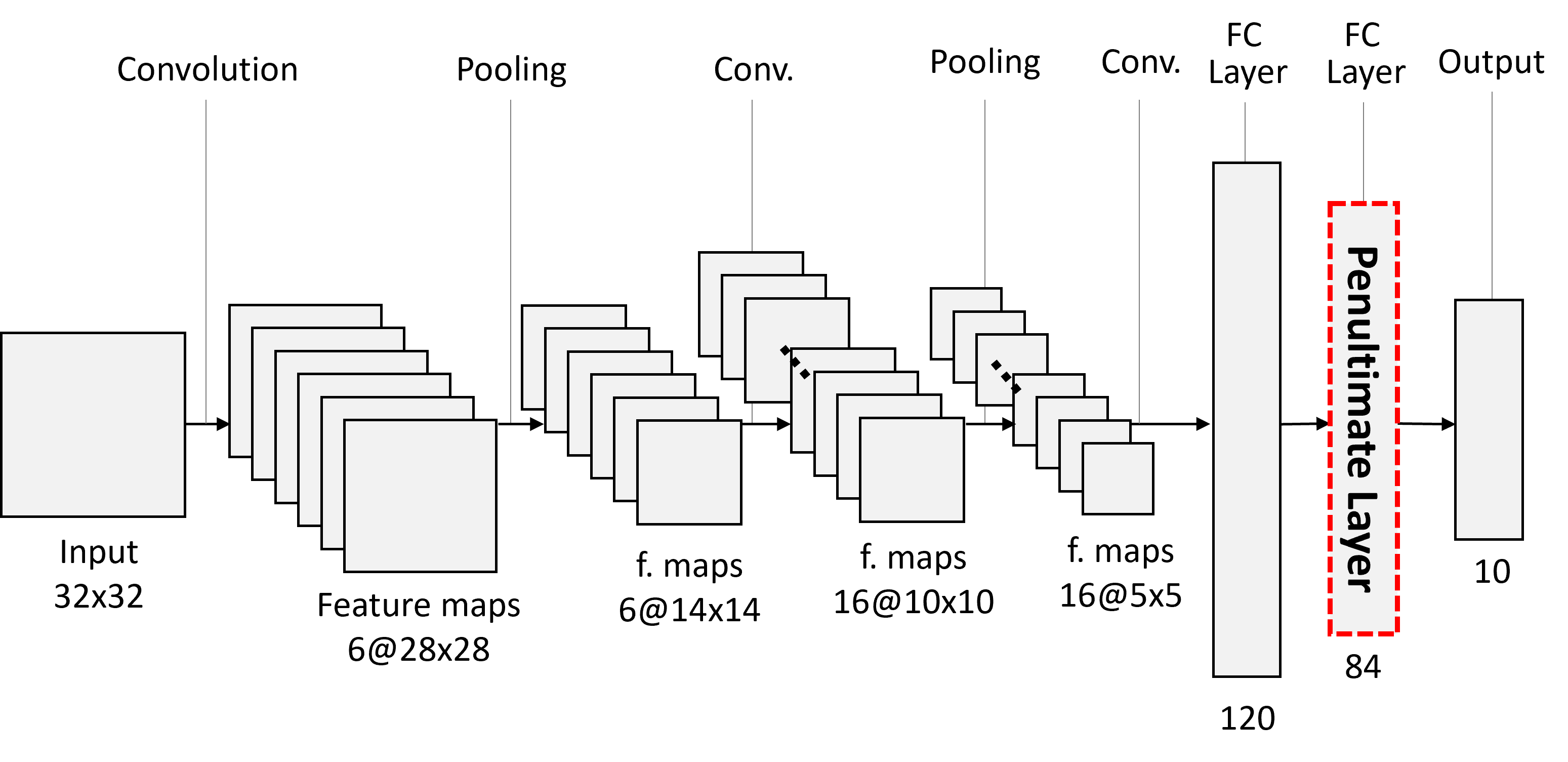}
    \caption{LeNet (CNN) architecture and Penultimate Layer}
    \label{fig:plr}
    \vspace{-5mm}
\end{figure}

The penultimate layer is a neural network model's second-to-last layer (i.e., before the softmax layer), as shown in \autoref{fig:plr}. Wang et al. \cite{wang2022flare} discovered that PLRs are effective in distinguishing malicious models from benign ones; within benign clients, PLRs follow the same distribution, whereas malicious PLRs do not—across datasets and neural network architectures. Specifically, they demonstrated that distances among benign PLRs are smaller than those between benign and malicious PLRs. Their proposed method, FLARE, assigned a trust score to each client based on pairwise PLR discrepancies defined by Maximum Mean Discrepancy among all model updates, allocating lower values to those farther from the benign distribution. The model updates are then scaled and averaged, weighted by each client's trust score. Meanwhile, FLARE continuously redirected the global model using server-owned raw data, increasing the risk of data leakage and jeopardizing data privacy.

\subsection{Centered Kernel Alignment}
CKA \cite{kornblith2019similarity} is a highly accurate similarity metric used to measure how similar two differently initialized or trained neural networks are, producing a value between 0 (no similarity) and 1 (identical). It is computed as:
\[CKA(X,Y) = \frac{HSIC(X,Y)}{\sqrt{HSIC(X,X)HSIC(Y,Y)}} 
    = \frac{{\lVert K(X^{T})K(Y) \rVert}^{2}}{{\lVert K(X^{T})K(X) \rVert}{\lVert K(Y^{T})K(Y) \rVert}}
\]
where $HSIC$ refers to the Hilbert-Schmidt Independence Criterion \cite{gretton2005measuring}, a non-normalized variant of CKA, and $K$ denotes the RBF kernel. CKA constructs similarity kernel matrices in the weight space and compares them to characterize the representation space, enabling comparison across layers with different widths and initialization schemes.

CKA satisfies several key properties desirable in neural network similarity metrics. It is non-invariant to invertible linear transformations, meaning similarity scores change if such transformations are applied. This property is vital; the gradient descent algorithm is not invariant to these transformations because similarity metrics invariant to linear transformations are inaccurate on models trained with gradient descent. It is also invariant to orthogonal transformations and isotropic scaling, both of which are desirable for neural networks trained with gradient descent owing to their stochastic nature. These properties make CKA a precise and reliable metric for measuring similarities between neural networks, particularly when comparing models potentially generated by malicious clients.

CKA has been utilized to address data heterogeneity issues, taking into account that the similarity of non-IID data is notably lower than that of IID data \cite{vahidian2022rethinking}, especially in specific network layers \cite{son2022comparisons}. 
However, it has not yet been explored as a defense mechanism against malicious clients, where model similarity could reveal suspicious deviations under non-IID distributions.

\section{Threat Model} \label{Section3}
\subsection{Attackers' Goals}
We consider two primary attack strategies in the context of FL: untargeted model poisoning and targeted backdoor attacks. Untargeted attacks (Fang attacks \cite{fang2020local}) aimed to evade robust aggregation rules like Krum and Coordinate-wise Median or Trimmed Mean. Attackers manipulate the model parameters to degrade the global model's overall accuracy. 
Targeted backdoor attacks, represented by Bhagoji et al. \cite{bhagoji2019analyzing} and DBA \cite{xie2020dba}, aim to deceive the global model into misclassifying specific data samples the attacker chose. The objective is to assign a different label chosen by the attacker while maintaining high accuracy for the remaining classes, making the attack more inconspicuous.

\subsection{Attackers' Capability}
\begin{itemize} 
    \item An attacker on the client side can control multiple compromised clients. 
    \item An attacker has full control over at most $k < n/2$ clients.
    \item An attacker has knowledge about compromised clients' data, such as the current local model, previous global model, and hyperparameters.
    \item An attacker has no knowledge or control over the server and honest clients.
\end{itemize}
% , such as learning rate, optimization method, loss function, batch size, and the number of local epochs.
\subsection{Attack Strategy}

\subsubsection{Untargeted Model Poisoning Attacks \cite{fang2020local}} specifically aim to break Krum and Coordinate-wise Median (Coomed) aggregation rules, known to be byzantine failure tolerant. Their primary objective is to disrupt the model training process, thereby diminishing global test accuracy. The attacks involve manipulating model parameters, such as flipping the sign of malicious parameters, to steer the model in the opposite direction from its uninterrupted trajectory. Specifically, we denote the attack against Krum as \textbf{Untargeted-Krum} and the attack against Coomed as \textbf{Untargeted-Med}. 

\textit{Untargeted-Krum} \cite{fang2020local} alters malicious parameters resembling benign ones to maximize the chances of being selected by Krum. Specifically, the optimization problem is to find the maximum $\lambda, s.t.$ $w_{1} = Krum(w_{1}, ...,w_{m}, w_{m+1}, ..., w_{n}), w_{1} = w_{G}-\lambda s, w_{i} = w_{1},$ \text{for} $i = 2, 3, ..., m $
% \end{align*}
where $m$ is the number of attackers, $n$ is the total number of selected clients, $w_{G}$ is the previous global model, and $s$ is the sign of the global model parameter with no attack.
The upper bound of the $\lambda$ can be solved as follows:
% \begin{align*}
$\lambda \leq \frac{1}{(n-2m-1)\sqrt{d}} \cdot \min_{m+1 \leq i \leq n} \sum_{l \in \Gamma_{w_{i}}^{n-m-2}} D(w_{l}, w_{i}) + \frac{1}{\sqrt{d}} \cdot \max_{m+1\leq i \leq n}D(w_{i}, w_{G})$
% \end{align*}
where $d$ is the number of parameters in the global model, $\Gamma$ is the set of $n-m-2$ benign local models having the smallest Euclidean distance to $w_{i}$, and $D$ is the Euclidean distance. 
$\lambda$ is halved until one of the compromised models is selected, or $\lambda$ is less than a threshold. 

\textit{Untargeted-Med} \cite{fang2020local} manipulates the model parameters based on the maximum and minimum so that chosen coordinate-wise median values direct toward an inverse direction. 
The attack starts with defining the maximum and minimum of the $j$th local model parameters on the benign clients, $w_{max,j} = \text{max}\{w_{(m+1),j}, w_{(m+2),j}, ..., w_{n,j}\}$ and $w_{min,j} = \text{min}\{w_{(m+1),j}, w_{(m+2),j}, ..., w_{n,j}\}$, respectively.
Also, to avoid sampled $m$ numbers being outliers, if $s_{j}=-1$, $m$ numbers in [$w_{max,j}, b\cdot w_{max,j}$] (when $w_{max,j} > 0$) or [$w_{max,j}, b/w_{max,j}$] (when $w_{max,j} \leq 0$) are randomly sampled. Otherwise, $m$ numbers [$w_{min,j}/b, w_{min,j}$] (when $w_{min,j} > 0$) or [$b\cdot w_{min,j}, w_{min,j}$] (when $w_{min,j} \leq 0$) are randomly sampled. We set $b=2$ as the same as the paper \cite{fang2020local}.  

\subsubsection{Targeted Backdoor Attacks} are based on \textbf{\cite{bhagoji2019analyzing}}. Each malicious client owns one sample of mislabeled images and trains the local model on it. We trained the model on backdoor tasks along with the main task(s); the training went on for both malicious and benign tasks to maintain benign accuracy such that backdoor training remains stealthy. Then we boosted malicious clients' updates to negate the combined effect of the benign agent: $w_{i}^{t} = w_{G}^{t-1} + \alpha_{m}(w_{i}^{t-1}-w_{G}^{t-1}) \text{ for } i = 1, ..., m$
where $t$ is the current epoch and $\alpha_{m}$ is a boosting factor. 
For \textbf{DBA} \cite{xie2020dba}, we implant split backdoor triggers into the input so that the global model can be attacked in a distributed manner. For more details, refer to \cite{xie2020dba}. 
In CIFAR10, for example, we manipulate the model parameters to misclassify an image of `airplanes' as `birds', whereas normally it would classify the image as other classes.

\begin{figure*}
    \centering
    \includegraphics[width=\linewidth]{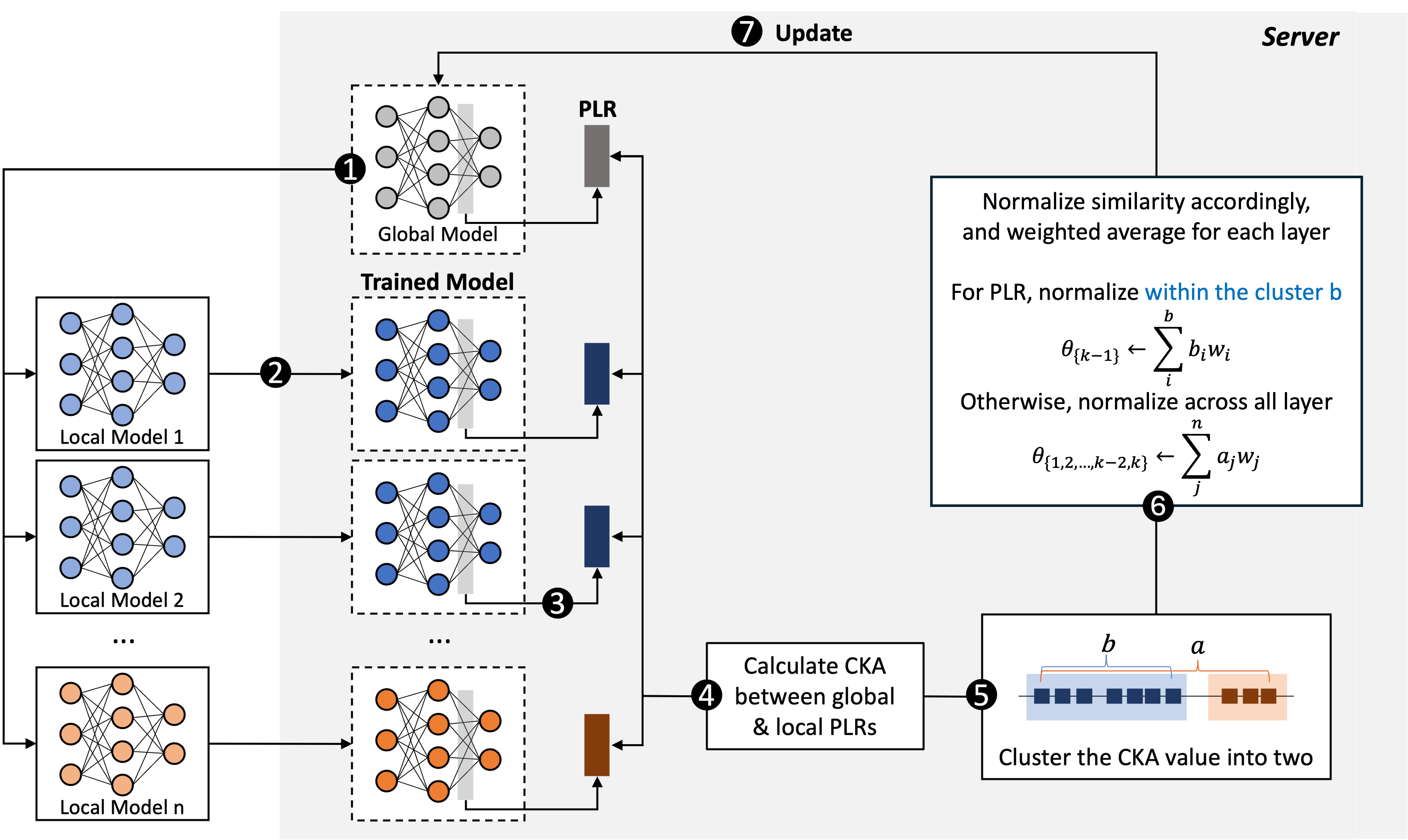}
    \caption{An Overview of FedCC. $k$, $n$, $a$, and $b$ indicate the kth layer, the number of participating clients, normalized similarity across all layers (i.e., \texttt{across\_cka}), and normalized similarity within the cluster (i.e., \texttt{within\_cka}), respectively. }
    \label{fig:overview}
    \vspace{-5mm}
\end{figure*}

\section{FedCC: Robust Aggregation against Poisoning Attacks} \label{Section4}

\subsection{Overview of FedCC}
\autoref{fig:overview} depicts an overview of FedCC. 
(1) The server initializes and broadcasts the global model to $n$ clients, selected from a total of $k$ candidates with selection fraction $C$.
(2) Each client trains a local model on its private dataset and sends the updated weights to the server—this is where our proposed method intervenes.
(3) The server extracts PLRs from both the global model and each local model, and 
(4) compares RBF Kernel CKA values between them. 
(5) The CKA values are then clustered into two, potentially representing benign and malicious clients. 
(6) CKA similarity values are normalized across all clients for each layer as $a$. For the second-to-last layer (PLR), normalization is performed within the larger cluster as $b$, under the assumption that malicious clients are fewer than $n/2 - 1$.
(7) Finally, the server aggregates each layer of local models weighted by $a$, but the PLR weighted by $b$, to update a global model for the current epoch and proceeds to the next epoch by distributing the updated global model to the newly selected $n$ clients. 

\subsection{Detailed Design}

A key challenge in filtering malicious clients under non-IID conditions is the server's ignorance of the underlying data distribution.
Even if the data is benign, local learning models can exhibit significant angular or magnitude differences if they are non-IID. Thus, we require a similarity metric independent of data distribution and not influenced by the distance or direction of model parameters.

\subsubsection{Penultimate Layer} is the output of the second-to-last layer before the softmax layer in CNN. In federated learning, where the server cannot access local training data, it receives trained model parameters (weights) from selected clients. 
Thus, the received weights serve as the sole basis for detecting malicious behavior.
The PLR captures the final representation produced by convolution and pooling before reaching a classification decision. 
We hypothesize that the PLR contains the most task-relevant, discriminative features.

\begin{wrapfigure}{l}{0.5\textwidth}    
\centering
    \includegraphics[width=\linewidth]{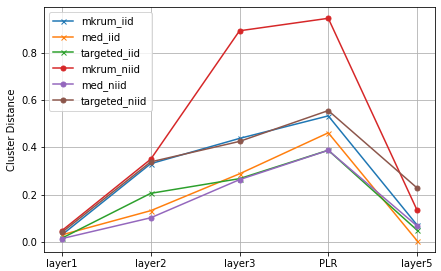}
    \caption{Cluster Distance of Each Layer}
    \label{fig:cluster_distance}
    \vspace{-5mm}
\end{wrapfigure}

To test this, we measure the distance between two clusters of clients (benign and malicious) using dendrograms with a single linkage metric and correlation distance method. \autoref{fig:cluster_distance} presents cluster distances of each layer under different attack scenarios, such as untargeted-mKrum attacks in IID or non-IID settings. Observably, the cluster distance is the most considerable in PLR to other layers, indicating the highest degree of differences. Based on this observation, we conclude that PLR contains the most distinguishable and indicative information. Therefore, we utilize the weights of the penultimate layer to compare their similarity to the global model.  

\begin{table}
% \vspace{-5mm}
\caption{Comparison of Performance with Various Similarity Metrics}
\begin{center}
\setlength\tabcolsep{2.8pt}
\begin{tabular}{ccccccc}
\toprule
Method & \multicolumn{2}{c}{Fang-Med} & \multicolumn{2}{c}{Fang-mKrum}&\multicolumn{2}{c}{Targeted} \\ \cmidrule(lr){2-3}\cmidrule(lr){4-5}\cmidrule(lr){6-7}
& IID & NIID & IID & NIID & IID & NIID \\
\midrule
Kernel CKA& \textbf{69.20} & \textbf{41.00} & \textbf{70.22} & \textbf{43.24} & \textbf{71.44/6e-07} & \textbf{54.62/0.0118}  \\
Linear CKA& 10.00 & 13.13 & 64.09 & 39.55 & 71.02/0.0007 & 49.53/0.0616 \\
MMD       & 63.39 & 40.90 & 69.69 & 32.27 & 70.85/1e-09 & 50.51/9e-05 \\
Cosine    & 68.82 & 33.90 & 68.81 & 10.04 & 69.76/0.0002 & 53.66/0.0529 \\
Euclidean & 69.06 & 27.82 & 68.54 & 41.57 & 69.17/0.0221 & 52.20/0.0015 \\
\bottomrule
\end{tabular}
\label{tab:diffsim}
\end{center}
\vspace{-5mm}
\end{table}

\subsubsection{Kernel CKA\cite{kornblith2019similarity}} is a similarity metric to assess the similarity or dissimilarity between two neural networks. 
It enables the comparison of representations across layers or among models trained under varying conditions.
The sophisticated similarity comparison ability of Kernel CKA makes it suitable for our purpose. 

To justify the effectiveness of Kernel CKA, we compared its performance with other similarity metrics such as linear CKA, MMD (not normalized CKA), cosine similarity (angle), and Euclidean similarity (distance). We evaluated the test accuracy of FedCC using the CIFAR10 dataset under two untargeted attacks and one targeted attack, separately in IID and non-IID settings. The results are summarized in \autoref{tab:diffsim}. In the `targeted' column, the values represent the combination of test accuracy and backdoor confidence, with higher test accuracy and lower backdoor confidence indicating better performance. 
We observed that Kernel CKA consistently yields the highest performance across experiments; therefore, we adopted it to measure PLR similarity.

\subsubsection{FedCC} is an aggregation method that combines local clients' model parameters based on the Kernel CKA similarities between the global and local models' PLRs. The complete algorithm is provided in Algorithm~\autoref{algo}. The server first extracts the PLRs from both the global and local models, then computes the RBF Kernel CKA between them. We focus on PLRs specifically, as they exhibit the most distinguishable differences between benign and malicious models, as shown in earlier sections. The RBF Kernel CKA is chosen for its effectiveness in capturing nuanced similarity differences under non-IID and adversarial settings.
The resulting similarity values are clustered into two groups using a simple K-means algorithm---though any binary clustering method would suffice. Assuming that fewer than half of the clients are adversarial ($n/2$), the larger cluster is designated as the set of `candidates.' 
This approach does not require manual thresholding; the fixed $K=2$ clustering aligns with the practical assumption that fewer than half of the clients are adversarial, and empirically, this dynamic separation has shown consistent effectiveness across rounds and datasets.

Given that earlier CNN layers capture global features while later layers encode local ones \cite{krizhevsky2012imagenet, nichani2024provable, mundt2018rethinking}, we apply two types of normalization: \texttt{within\_cka} (denoted $b$ in \autoref{fig:overview}), computed within the dominant (larger) cluster, and \texttt{across\_cka} (denoted $a$), computed across all clients. For the PLR (second-to-last layer), we use \texttt{within\_cka}; for all other layers, we use \texttt{across\_cka} to compute layer-wise weighted averages. This helps preserve the scale of the aggregated parameters, as the weights are normalized to sum to one.

FedCC is also designed to remain effective against adaptive attacks, including distributed adaptive attacks like DBA. Unlike defenses based on outlier detection in parameter space, FedCC evaluates representational similarity using Kernel CKA between PLRs of local and global models. This enables the detection of semantic deviations even when malicious updates mimic benign gradients. By clustering clients based on CKA similarity and applying soft, layer-specific weighting---rather than hard rejection---FedCC attenuates the influence of suspicious clients without excluding them outright. This makes it robust against attacks that aim to blend in with benign behavior, particularly under non-IID conditions where traditional defenses often fail.

\paragraph{Theoretical Insight.}
While FedCC is primarily supported by empirical evidence, we briefly offer a theoretical perspective on its robustness. Let 
$\mathcal{R}_i$ denote the PLR of client $i$, where benign clients’ PLRs are drawn from distribution $\mathcal{D}_b$ and malicious clients’ from $\mathcal{D}_a$. If the inter-client Kernel CKA similarity within $\mathcal{D}_b$ is significantly higher than between $\mathcal{D}_b$ and $\mathcal{D}_a$, then clustering based on PLR similarity can effectively separate benign from malicious clients. Prior work has empirically demonstrated that PLRs from benign clients exhibit high intra-distribution similarity across diverse data distributions \cite{wang2022flare}. FedCC leverages this separability by softly weighting updates: clients close to the majority cluster receive higher aggregation weights, while those that deviate receive less influence. This approach mitigates adversarial updates without relying on hard rejection, and remains effective even when malicious updates mimic benign gradients. Therefore, under the mild assumption that benign clients maintain high intra-cluster PLR similarity and form the majority, FedCC is theoretically expected to preserve robustness across a range of attack scenarios.

\begin{algorithm}
\caption{FedCC}
\hspace*{\algorithmicindent} \textbf{Input:} global\_w, \textit{n} local\_w ($w_1$, ..., $w_n$) \\
\hspace*{\algorithmicindent} \textbf{Output:} agg\_w, larger\_cluster\_members
\hspace*{\algorithmicindent}{\color{blue}\Comment{Get PLRs of local models}}
\begin{algorithmic}[1]

\For{$i < n$} 
    \State {local\_plrs[$i$] $\gets$ local\_w[$i$][second\_last\_layer]} 
\EndFor

{\noindent\color{blue}\Comment{Extract global PLR}}
\State glob\_plr $\gets$ global\_w[second\_last\_layer] 

{\noindent\color{blue}\Comment{Apply kernel CKA to the plrs}}
\State cka[$i$] $\gets$ kernel\_CKA(glob\_plr, local\_plrs[$i$]) 

{\noindent\color{blue}\Comment{Normalize similarities globally}}
\State \texttt{across\_cka} $\gets$ \texttt{normalize}(cka) 

{\noindent\color{blue}\Comment{Apply Kmeans clustering algorithm}}
\State kmeans $\gets$ \texttt{kmeans}(n\_clusters=2, cka) 
\State labels $\gets$ kmeans.labels\_
\State count $\gets$ counter(labels)
\State larger\_cluster $\gets$ 1
\If {count[$0$] $>$ count[$1$]} 
    \State larger\_cluster $\gets$ 0
\EndIf

{\noindent\color{blue}\Comment{Identify larger cluster members}}
\State {larger\_cluster\_members = where(labels == larger\_cluster)}  

{\noindent\color{blue}\Comment{Normalize similarities within larger cluster}}
\State \texttt{within\_cka} $\gets$ \texttt{normalize}(cka[larger\_cluster\_members]) 

\State Initialize agg\_weights as zero tensor

{\noindent\color{blue}\Comment{Differently weigh weights per layer}}
\For{layer in model\_layers} 
    \If{layer is up to second-to-last layer} 
        \State agg\_weights[layer] $\gets$ weighted average using \texttt{across\_cka}
    \ElsIf{layer is second-to-last layer} 
        \State agg\_weights[layer] $\gets$ weighted average using \texttt{within\_cka}
    \Else
        \State agg\_weights[layer] $\gets$ weighted average using \texttt{across\_cka}
    \EndIf
\EndFor

\State \textbf{return} agg\_weights, larger\_cluster\_members
\end{algorithmic}
\label{algo}
\end{algorithm}

\section{Experiments} \label{Section5}

\subsection{Dataset and Model Architecture}
We use three benchmark vision datasets: Fashion-MNIST (fMNIST) \cite{xiao2017fashion}, CIFAR10, and CIFAR100 \cite{krizhevsky2009learning}. 
Considering that the end devices are normally incapable of handling heavy computation due to resource or communication constraints, we used lightweight CNN models for experiments. Specifically, we use 4-layer, 5-layer CNN for fMNIST and CIFAR10, and LeNet \cite{lecun1998gradient} for CIFAR100, as summarized in \autoref{tab:cnn_all}. 
% with two convolutions and two fully connected (FC) layers, incorporating dropout with probabilities of 0.25 and 0.5 before and after the FC layer, respectively. For CIFAR10, we utilized a 5-layer-CNN with three convolutions, each followed by a max-pool layer and two FC layers, applying dropout with a probability of 0.5 before the first FC layer. For CIFAR100, we adopted LeNet \cite{lecun1998gradient}. 

\begin{table}[hbt!]
\caption{CNN Architectures for fMNIST, CIFAR-10, and CIFAR-100}
\begin{center}
\setlength\tabcolsep{5pt}
\renewcommand{\arraystretch}{0.85}
\begin{tabular}{cccccc}
\toprule
Dataset     & Layer       & In    & Out   & Ker / Str / Pad    & Activation \\
\midrule
\multirow{7}{*}{fMNIST} 
            & conv2d\_1   & 1     & 64    & $5\times5$ / 1 / 0 & ReLU \\
            & conv2d\_2   & 64    & 64    & $5\times5$ / 1 / 0 & ReLU \\
            & dropout     & -     & -     & 0.25               & - \\
            & flatten     & -     & 25600 & -                  & - \\
            & fc\_1       & 25600 & 128   & -                  & - \\
            & dropout     & -     & -     & 0.5                & - \\
            & fc\_2       & 128   & 10    & -                  & - \\
\midrule
\multirow{10}{*}{CIFAR-10} 
            & conv2d\_1   & 3     & 64    & $3\times3$ / 1 / 0 & ReLU \\
            & maxpool2d   & -     & -     & $2\times2$ / - / 0 & - \\
            & conv2d\_2   & 64    & 64    & $3\times3$ / 1 / 0 & ReLU \\
            & maxpool2d   & -     & -     & $2\times2$ / - / 0 & - \\
            & conv2d\_3   & 64    & 64    & $3\times3$ / 1 / 0 & ReLU \\
            & maxpool2d   & -     & -     & $2\times2$ / - / 0 & - \\
            & flatten     & -     & 256   & -                  & - \\
            & dropout     & -     & -     & 0.5                & - \\
            & fc\_1       & 256   & 128   & -                  & - \\
            & fc\_2       & 128   & 10    & -                  & - \\
\midrule
\multirow{8}{*}{CIFAR-100} 
            & conv2d\_1   & 3     & 6     & $5\times5$ / 1 / 0 & ReLU \\
            & maxpool2d   & -     & -     & $2\times2$ / 2 / 0 & - \\
            & conv2d\_2   & 6     & 16    & $5\times5$ / 1 / 2 & ReLU \\
            & maxpool2d   & -     & -     & $2\times2$ / 2 / 0 & - \\
            & flatten     & -     & 44944 & -                  & - \\
            & fc\_1       & 44944 & 120   & -                  & ReLU \\
            & fc\_2       & 120   & 84    & -                  & ReLU \\
            & fc\_3       & 84    & 100   & -                  & - \\
\bottomrule
\end{tabular}
\label{tab:cnn_all}
\vspace{-5mm}
\end{center}
\end{table}

\subsection{Non-IID Simulation}
Non-IID is a more feasible assumption considering the diverse and massive nature of data in practice. To standardize non-IID settings, we use the Dirichlet distribution, which is widely adopted for modeling client-level label skew \cite{li2022federated}. The concentration parameter $\alpha$ controls the degree of heterogeneity; smaller values lead to clients holding examples from fewer classes. We set $\alpha = 0.2$ to simulate moderate heterogeneity, following prior works \cite{huang2024lockdown, zhang2022flip, wang2022flare}. This value is commonly used as it balances realism and trainability: smaller values (e.g., $\alpha < 0.1$) can induce excessive label imbalance and hinder convergence, while larger values (e.g., $\alpha > 0.5$) reduce the heterogeneity to near-IID conditions \cite{borazjani2025redefining}.

\subsection{Experimental Setup and Baselines}
We implement two untargeted attacks from \cite{fang2020local} and two targeted attacks from \cite{bhagoji2019analyzing} and \cite{xie2020dba}, along with defense baselines including Krum, multi-Krum \cite{blanchard2017machine}, Coomed \cite{yin2018byzantine}, Bulyan \cite{guerraoui2018hidden}, FLTrust \cite{cao2020fltrust}, FLARE \cite{wang2022flare}, and FedCC. We use ten clients with a 1.0 client participation fraction unless stated otherwise. As in \cite{wang2022flare}, we have three adversaries for each untargeted attack and one for the targeted attack. In DBA, we randomly select four malicious clients and implant five distributed triggers per batch. To compensate for delayed attack effects, we use FedAvg until epoch 30, after which each defense method is applied. Note that, for a fair comparison, FLTrust is computed using its previous global model without auxiliary data. Benign clients undergo three local epochs, while compromised clients undergo six, with training done using Adam optimizer (learning rate = 0.001).

\subsection{Evaluation Metrics}
We evaluated our defense mechanism using two metrics: backdoor confidence (\textit{confidence}) and global model test accuracy (\textit{accuracy}). The confidence metric measures misclassification likelihood, ranging from 0 to 1 (lower values are better). Accuracy reflects the global model's overall test performance under defense methods during attacks, ranging from 0 to 100 (higher values are better). Both metrics were reported under backdoor attacks, with a focus on accuracy for untargeted attacks, as the attacker's goal is to reduce overall accuracy.

\section{Results and Discussions} \label{Section6}
In this section, we report the performance of our FedCC against two untargeted and two targeted attacks in both IID and non-IID environments. 
We additionally demonstrate the defense effect against a distributed adaptive attack, DBA \cite{xie2020dba}. 
Unlike centralized backdoor attacks, the DBA splits a trigger pattern across multiple adversaries, evading detection by anomaly detection. We show the result of CIFAR10, with various defense methods applied.

\subsection{Non-IID Data Environment}

Non-IID is a more feasible assumption considering the diverse and massive nature of data in practice. To standardize non-IID settings, we employ Dirichlet distribution that 0 indicates the most heterogeneity, and 100 mimics homogeneity \cite{lin2020ensemble}. A concentration parameter $\alpha$ controls the degree of non-IID; the smaller $\alpha$ is, the more likely the clients hold examples from only one randomly chosen class. Since $\alpha$ being 0.2 represents a highly non-IID scenario based on \cite{huang2024lockdown, zhang2022flip}, we set it as such, following \cite{zhang2022flip}.

\subsubsection{Untargeted Model Poisoning Attacks in Non-IID Setting}

\autoref{tab:untargeted_niid_performance} provides the global model's accuracy across three datasets under two untargeted (Fang-Krum and Fang-Med) in a non-IID environment.
A notable finding is that robust aggregation algorithms often yield lower accuracy than simple averaging (fedAvg). This discrepancy arises from the imperfect identification of malicious clients, leading to the erroneous aggregation of their weights along with those of benign clients.
Under Fang-Krum, Krum, multi-Krum, and Bulyan experience significant accuracy drops (16.51\%, 45.88\%, and 13.39\%, respectively, compared to FedAvg's 57.14\%), reflecting the attack's design to degrade Krum-based defenses. 
Coomed also struggles, as median weights fail to represent benign clients effectively in non-IID settings. FLTrust mitigates Fang-Krum but performs better on CIFAR10 than fMNIST, likely due to fMNIST's simpler gradients and lower variance, which challenge its ability to distinguish malicious updates.
FLARE's lower accuracy stems from its approach of scaling the entire model weights uniformly, which ignores knowledge contributed by individual clients. In contrast, FedCC, which selectively averages weights layer-wise, achieves superior performance by boosting or minimizing weights based on their alignment with benign updates. FedCC demonstrates the highest accuracy (71.13\%, 52.06\%, and 14.51\% for fMNIST, CIFAR10, and CIFAR100, respectively) and shows significant improvement in CIFAR100 experiments.

Turning to the Fang-Med attack, FedAvg experiences a significant drop in accuracy due to the attack's creation of outliers by deviating from median values. Consequently, Coomed and Bulyan perform relatively well since their coordinate-wise median and trimmed mean methods disregard outliers effectively.
FLARE continues to perform poorly for a similar reason mentioned earlier. Between the Krum-based methods, multi-Krum and Coomed are more robust than Krum, as they aggregate multiple local models, providing better resilience to outliers. Similarly, Bulyan's averaging of multiple clients enhances robustness but falls short of Coomed, particularly in non-IID environments.
Finally, FedCC achieves the highest accuracy (72.76\%, 47.85\%, and 16.12\% for each dataset) without sharing raw data, showcasing its exceptional performance. This success is attributed to two factors: (1) highly distinguishable information within PLRs and (2) higher CKA similarity between benign clients than between benign and malicious clients.

\begin{table}
\vspace{-5mm}
\footnotesize
\setlength{\tabcolsep}{0.75pt} % Reduce column padding
\caption{Test Accuracy under untargeted attacks in Non-IID setting.}
\begin{center}
\begin{tabular}{c|c|cccccccc}
\toprule
\textbf{Case}&\textbf{data} & \textbf{FedAvg} &\textbf{Krum} & \textbf{MKrum} &\textbf{Coomed} &\textbf{Bulyan} & \textbf{FLTrust} &\textbf{FLARE} &\textbf{FedCC}\\ \hline
\multirow{3}{*}{\shortstack{\textbf{Fang}\\ \textbf{-Krum} \\ non-IID}}
&fM  & 57.14 & 16.51 & 45.88 & 57.12 & 13.39 & 60.8 & 49.54 & \textbf{71.13}  \\
&C10 & 33.69 & 15.38 & 20.5 & 35.7 & 19.23 & 41.87 & 17.03 & \textbf{52.06} \\
&C100& 2.27 & 1       & 4.95 & 7.85 & 0.98 & 11.04 & 7.46 & \textbf{14.51} \\ \hline
\multirow{3}{*}{\shortstack{\textbf{Fang} \\ \textbf{-Med} \\ non-IID}}
&fM  & 16.32 & 49.33 & 66.84 & 68.9 & 64.12 & 18.96 & 52.25 & \textbf{72.76} \\
&C10 & 10.02 & 25.06 & 45.44 & 40.23 & 32.47 & 10 & 14.59 & \textbf{47.85} \\
&C100& 1 & 6.24 & 14.52 & 10.27 & 6.91 & 1.09 & 1 & \textbf{16.12} \\
\bottomrule
\end{tabular}
\label{tab:untargeted_niid_performance} 
\end{center}
\vspace{-7mm} 
\end{table}

\subsubsection{Targeted Backdoor Attacks in Non-IID Setting}
\begin{figure}
    % \vspace{-5mm}
    \centering
    \includegraphics[width=\linewidth]{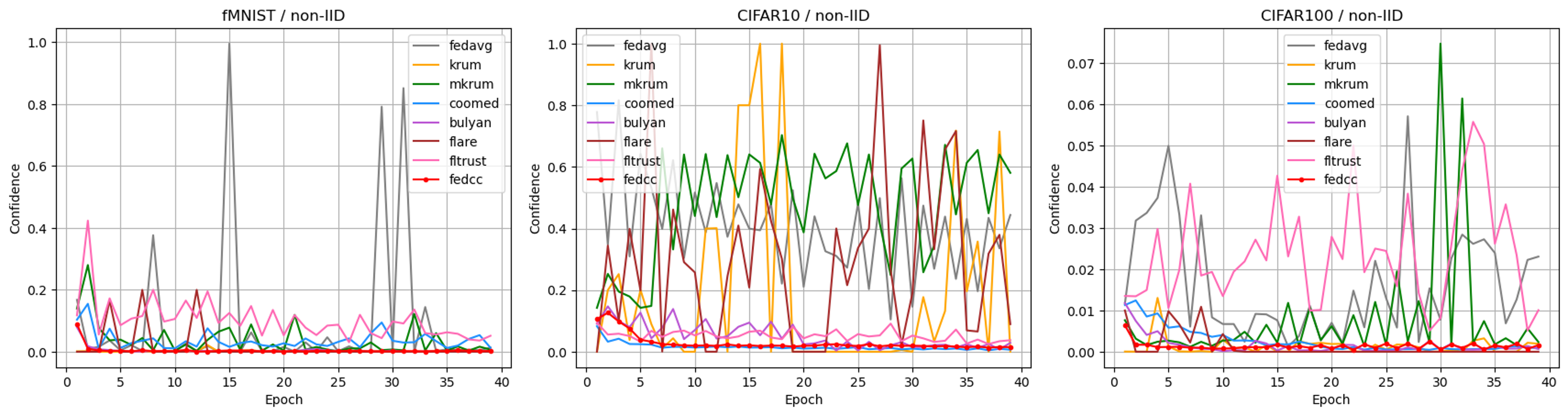}
    \caption{Confidence of Backdoor Task for targeted attacks in Non-IID settings.}
    \label{fig:confidence_niid}
    \vspace{-5mm}
\end{figure}

\autoref{tab:targeted_niid_performance} summarizes test accuracy under targeted attacks in a non-IID setting. To further illustrate the impact, \autoref{fig:confidence_niid} provides a visual representation of backdoor confidence. It is evident that Krum, which selects a single client's weights as the global model, shows reduced robustness compared to methods that aggregate weights from multiple clients. In contrast, Coomed, multi-Krum, and Bulyan exhibit higher test accuracy.

FLARE reduces main task accuracy for reasons similar to those observed in previous experiments. FedCC achieves the highest test accuracy. Since targeted attacks aim to misclassify a specific target class while maintaining the main task accuracy, the best performance of FedCC is attributed to leveraging knowledge from earlier stages of training. This result highlights the superiority of using sophisticated weighted knowledge and filtering malicious clients via CKA similarity, which outperforms first-order statistical methods like mean or median.

A similar trend is observed with the DBA attack, except for multi-Krum; the lowest accuracy occurs because all four malicious clients, each with a distributed trigger, are treated as outliers and excluded from aggregation. In contrast, FedCC preserves high main task accuracy (52.28\%) even under DBA, outperforming all baselines. This result highlights its robustness against adaptive strategies that distribute triggers across multiple compromised clients—a setting specifically designed to evade anomaly-based filtering.

\begin{table}
\vspace{-5mm}
\footnotesize
\setlength{\tabcolsep}{0.75pt} % Reduce column padding
\caption{Test Accuracy under targeted attacks in Non-IID setting.}
\begin{center}
\begin{tabular}{c|c|cccccccc}
\toprule
\textbf{Case}&\textbf{data} & \textbf{FedAvg} &\textbf{Krum} & \textbf{MKrum} &\textbf{Coomed} &\textbf{Bulyan} & \textbf{FLTrust} &\textbf{FLARE} &\textbf{FedCC}\\ \hline
\multirow{3}{*}{\shortstack{\textbf{Target} \\ non-IID}}
&fM  & 75.65 & 45.27 & 65.97 & 71.70 & 57.96 & 61.82 & 64.31 & \textbf{75.66} \\
&C10 & 36.16 & 14.98 & 30.72 & 48.97 & 40.11 & 44.06 & 10.18 & \textbf{51.56} \\ %52.60 
&C100& 4.46 & 6.18 & 6.90 & 12.04 & 11.16 & 12.95 & 1.14 & \textbf{15.26}  \\ \hline
\textbf{DBA} & C10 & 38.56 & 24.94 & 7.09 & 44.45 & 34.19 & 51.49 & 38.73 & \textbf{52.28} \\
\bottomrule
\end{tabular}
\label{tab:targeted_niid_performance}
\end{center}
\vspace{-5mm}
\end{table}

Notably, \autoref{fig:confidence_niid} demonstrates that FedCC significantly reduces backdoor confidence. Note that since the backdoor confidence of DBA fluctuates a lot due to its distributed nature, we omit it for brevity. Unlike other methods, such as FLTrust or multi-Krum, which exhibit fluctuating confidence values, FedCC maintains consistently low confidence, emphasizing its resilience and independence from client-specific data distributions. This advantage stems from the effective utilization of CKA, enabling similarity calculations even when models are trained on different datasets with varying distributions. While other baselines, like Bulyan or FLTrust, also reduce confidence, FedCC not only mitigates the backdoor attack but also sustains---or improves---accuracy by precisely identifying and down-weighting malicious clients.

\paragraph{Summary of Non-IID Robustness.} Across all attack types, FedCC consistently demonstrates superior performance in non-IID environments---outperforming prior defenses in both main task accuracy and backdoor confidence. This robustness stems from the use of Kernel CKA over penultimate layer representations, which captures distributional structure even under client heterogeneity. Unlike methods based on geometric or statistical outlier filtering, FedCC’s representation-based, layer-aware weighting enables it to preserve benign signals while attenuating subtle or distributed malicious updates, making it particularly effective under realistic non-IID conditions.

\subsection{IID Data Environment}
In an IID setting, we evenly divide the training dataset among all clients so that each client's data distribution is identical and of the same size.   

\subsubsection{Untargeted Model Poisoning Attacks in IID Setting}

\autoref{tab:untargeted_iid_performance} shows the test accuracy of the global model trained on three datasets under untargeted attacks in IID settings. While trends in these experiments are similar to those in the non-IID setting, the accuracy values are slightly higher due to the more consistent data distribution across and updates from clients in the IID case.
A notable deviation from the non-IID experiments is when Coomed is applied: the accuracy increases from 75.55\% (FedAvg) to 87.62\% (Coomed). This indicates that the coordinate-wise median values closely align with the global model's weights with minimal deviation and represent benign models well, as the adversary's impact is limited due to the majority of benign clients. 
FedCC achieves the highest accuracy across all methods (89.57\%, 69.84\%, and 18.47\% for each dataset) due to its effective CKA similarity measurement and sophisticated layer-wise aggregation strategy.

Under the Fang-Med attack, the accuracy of FedAvg experiences a significant drop (89.89\% to 20.86\%) due to the peculiarity of outliers when the clients have IID data. Similar to the result of Fang-Krum, Krum, Coomed, and Bulyan demonstrate similar accuracy values, indicating that coordinate-wise median values indeed represent benign clients' parameters. 
FLTrust, however, is less effective in mitigating the Fang-Med attack than Fang-Krum, even in the IID setting; it then means that the performance difference is not due to data distribution but rather its inherent vulnerability. In Fang-Krum, malicious updates are deliberately far from the global model, allowing FLTrust to identify discrepancies easily. In contrast, the malicious updates in Fang-Med are subtler and only slightly deviate from the global model or median, making it harder for FLTrust to distinguish them from benign updates.
Meanwhile, FedCC consistently achieves the highest accuracy (89.66\%, 70.52\%, and 17.83\% for each dataset), demonstrating its capability to mitigate attacks involving more nuanced changes to model updates without extreme outliers.

\begin{table}
\vspace{-5mm}
\footnotesize
\setlength{\tabcolsep}{0.75pt} % Reduce column padding
\caption{Test Accuracy under untargeted attacks in IID setting.}
\begin{center}
\begin{tabular}{c|c|cccccccc}
\toprule
\textbf{Case}&\textbf{data} & \textbf{FedAvg} &\textbf{Krum} & \textbf{MKrum} &\textbf{Coomed} &\textbf{Bulyan} & \textbf{FLTrust} &\textbf{FLARE} &\textbf{FedCC}\\ \hline
\multirow{3}{*}{\shortstack{\textbf{Fang}\\ \textbf{-Krum} \\ IID}}
&fM  & 75.55 & 31.66 & 87.78 & 87.62 & 50.30 & 89.53 & 79.16  & \textbf{89.57} \\ 
&C10 & 49.67 & 40.86 &  63.42 & 57.40 & 12.67 & 68.25 & 25.77 & \textbf{69.84} \\ 
&C100& 13.72 & 1.04  & 7.64 & 6.17  & 1.59  &  17.14 & 7.49  & \textbf{18.47} \\ \hline
\multirow{3}{*}{\shortstack{\textbf{Fang} \\ \textbf{-Med} \\ IID}}
&fM  & 20.86 & 85.33 & 89.53 & 86.70 & 87.45 & 21.36& 71.08 & \textbf{89.66}\\ 
&C10 & 9.51  & 54.28 & 69.68 & 59.20 & 57.69 &  9.92 & 49.82 & \textbf{70.52} \\ 
&C100& 0.87  & 12.27 & 16.52 & 14.43 & 12.56 & 1.16 & 5.83  & \textbf{17.83} \\ 
\bottomrule
\end{tabular}
\label{tab:untargeted_iid_performance} 
\end{center}
\vspace{-7mm}
\end{table}

\subsubsection{Targeted Backdoor Attacks in IID Setting}
\begin{figure}
\vspace{-5mm}
\centering
    \includegraphics[width=\linewidth]{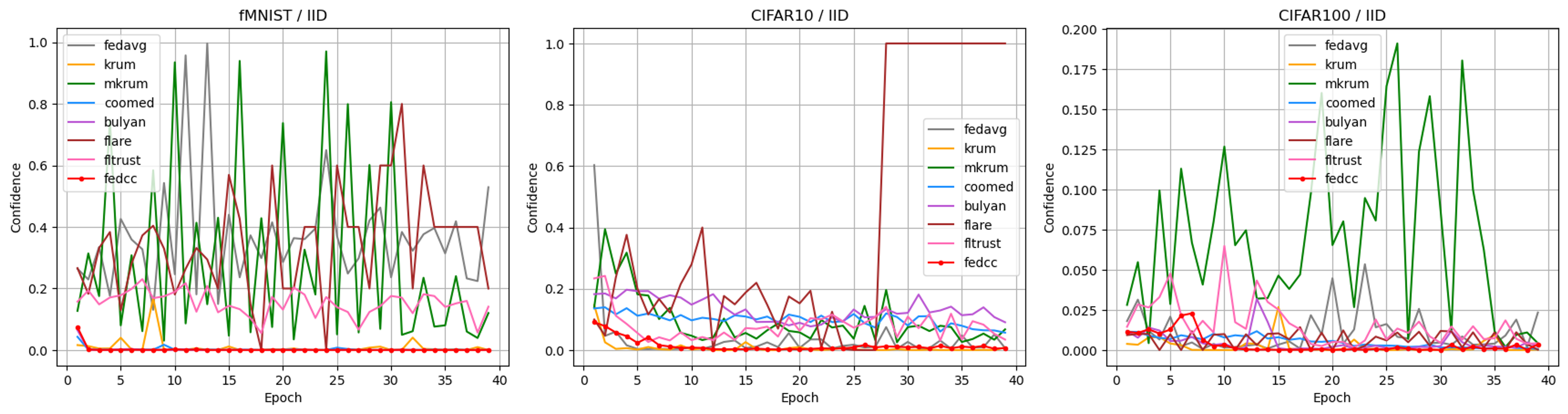}
    \caption{Confidence of Backdoor Task for targeted attacks in IID settings.}
    \label{fig:confidence_iid}
\vspace{-5mm}
\end{figure}

\autoref{tab:targeted_iid_performance} and \autoref{fig:confidence_iid} present the test accuracy and backdoor confidence under targeted backdoor attacks in IID settings. Similar to the untargeted attack scenario, the general test accuracy in IID settings is higher than in non-IID settings.

Reducing backdoor confidence is as critical as maintaining main task accuracy, as attackers aim to stealthily embed backdoor tasks. 
A key observation is FedCC's remarkable ability to reduce backdoor confidence to near zero while preserving high main task accuracy. In contrast, fluctuating backdoor confidence observed in Multi-Krum and Flare is largely due to the distributed trigger, which confuses their defense mechanisms and hampers their ability to filter outliers effectively. Other methods, such as Multi-Krum, Flare, and FLTrust, show significant variability in backdoor confidence, further highlighting FedCC's superiority.

Notably, FedCC achieved the highest main task accuracy across both targeted backdoor attacks and DBA, consistently neutralizing the backdoor task. These results underscore FedCC's ability to mitigate targeted backdoor attacks through precise and efficient integration of prior knowledge to preserve main task performance while zeroing out the attack confidence.

\begin{table}
\vspace{-5mm}
\footnotesize
\setlength{\tabcolsep}{0.75pt} % Reduce column padding
\caption{Test Accuracy under targeted attacks in IID setting.}
\begin{center}
\begin{tabular}{c|c|cccccccc}
\toprule
\textbf{Case}&\textbf{data} & \textbf{FedAvg} &\textbf{Krum} & \textbf{MKrum} &\textbf{Coomed} &\textbf{Bulyan} & \textbf{FLTrust} &\textbf{FLARE} &\textbf{FedCC}\\ \hline
\multirow{3}{*}{\shortstack{\textbf{Target} \\ IID}}
&fM  & 88.27 & 86.63 & 87.03 & 89.41 & 89.45 & 89.59 & 75.29 & \textbf{90.01} \\
&C10 & 64.68 & 57.69 & 71.19 & 69.85 & 68.76 & 68.61 & 11.09 & \textbf{71.64} \\ 
&C100& 13.83 & 5.72 & 17.57 & 13.24 & 15.11 & 17.08 & 1.03 & \textbf{18.61} \\ \hline
\textbf{DBA} & C10 & 10.00  & 35.58 & 51.40 & 10.00  & 16.16 & 56.69 & 10.00 & \textbf{58.04}\\
\bottomrule
\end{tabular}
\label{tab:targeted_iid_performance} 
\end{center}
\vspace{-5mm}
\end{table}

\subsection{Performance in Varying FL Settings}
We additionally measured the effectiveness of FedCC in various non-IID settings, including different numbers of malicious clients, fractions of participation, and numbers of local epochs. Since Coomed was as effective as ours, and FLARE also used PLRs, we compared three methods to FedCC, including FedAvg. 

\subsubsection{Impact of the Number of Malicious Clients}

We assessed the impact of different numbers of malicious clients under untargeted attacks, specifically Untargeted-Krum and Untargeted-Med. In this set of experiments, we kept the total number of clients and participation fraction fixed at 10 and 1, respectively. To adhere to the assumption that the number of attackers is less than half of the participating clients, we investigated the test accuracy considering scenarios involving up to four malicious clients.

\begin{figure}
% \vspace{-5mm}
    \centering
    \includegraphics[width=\linewidth]{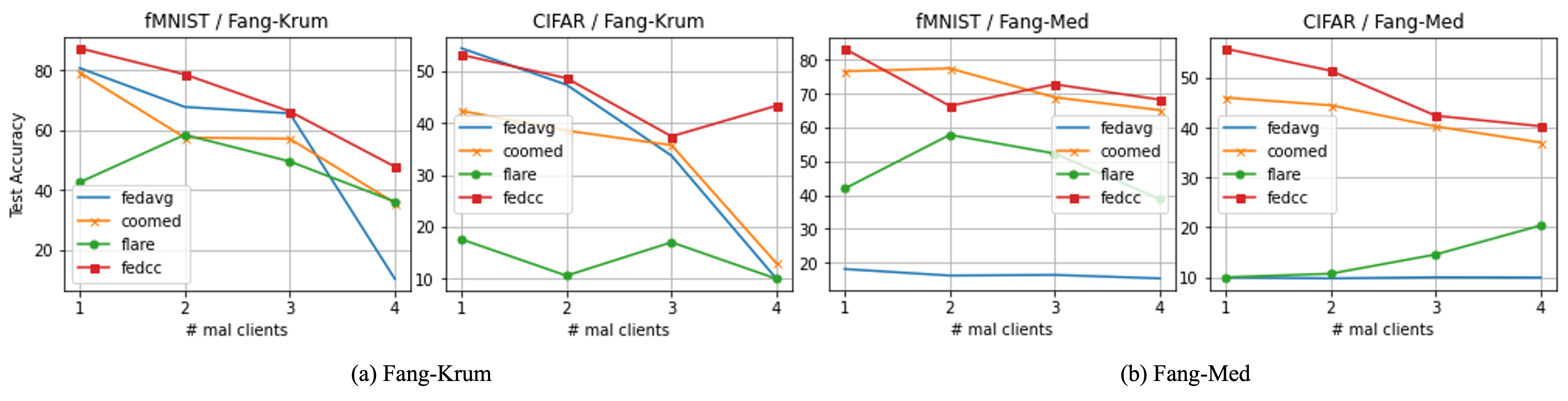}
    \caption{Test Accuracy with different numbers of malicious clients}
    \label{fig:diff_mal_clients}
\vspace{-5mm}
\end{figure}

\autoref{fig:diff_mal_clients} illustrates the accuracy in the given experimental settings, and it is evident that FedCC outperforms the other defense methods. We observe a general trend where the accuracy of defense methods decreases as the number of attackers increases. Notably, Coomed experiences the most significant drop in accuracy when transitioning from one malicious client to four malicious clients. This can be attributed to the fact that as the participation of malicious clients increases, there is a higher likelihood of the median values being influenced by their malicious contributions.

In contrast, FedCC does not rely solely on geometric measures such as angles or distances but instead leverages hidden correlations between networks to identify and filter out malicious clients. As a result, FedCC demonstrates superior accuracy in filtering out malicious clients, regardless of the number of attackers involved. It is important to note that the accuracy degradation observed as the number of malicious clients increases primarily due to aggregating fewer clients rather than the malicious clients themselves being selected.
These findings underscore the robustness and effectiveness of FedCC in defending against untargeted attacks, as it consistently outperforms other defenses across varying numbers of malicious clients.

\subsubsection{Impact of Fraction}
We examine the impact of the participating fraction of clients where the number of total clients is fixed at 100. The fractions are 0.1, 0.3, and 0.5, such that the numbers of clients being aggregated are 10, 30, and 50, respectively. The proportion of malicious clients is fixed at 0.3, such that the malicious clients are 3, 9, and 15, respectively. \autoref{tab:untargeted_diff_fraction} summarizes the accuracy. We can observe that the accuracy tends to increase as the participation fraction is greater under untargeted-Krum attacks. It implies that the more client participation in a non-IID setting, the more accurate the global model is if the attacks are mitigated. Under untargeted-Med attacks, the accuracy is the highest with FedCC and second highest with Coomed. 
Throughout the experiment, FLARE prevented the global model from convergence, and the test accuracy fluctuated. The fluctuation is presumably due to the weight scaling while randomly selecting clients. Different from previous experiments, the server chooses a fraction of the client every round, meaning not-yet-trained models can be selected. In this environment, weight scaling leveraged by FLARE constantly improperly scales the local weights, causing divergence and fluctuation.   
These experimental results indicate that selectively choosing clients based on CKA similarity is more reliable than taking the median value of each coordinate in a non-IID setting. We also observed that as the participation fraction grows, the accuracy grows. This is due to the rising number of aggregated local models; the more models are aggregated, the more general global model is generated.

\begin{table}
\vspace{-5mm}
\caption{Test Accuracy under untargeted attacks with different fractions of clients being selected.}
\begin{center}
\setlength\tabcolsep{2.8pt}
\begin{tabular}{cccccccccc}
\toprule
&&\multicolumn{4}{c}{\textbf{Untargeted-Krum}} & \multicolumn{4}{c}{\textbf{Untargeted-Med}} \\ \cmidrule(lr){3-6}\cmidrule(lr){7-10} 
\textbf{Frac} & \textbf{Data} & \textbf{FedAvg} &\textbf{Med} &\textbf{FLARE} &\textbf{FedCC} & \textbf{FedAvg} &\textbf{Med} &\textbf{FLARE} &\textbf{FedCC}\\ 
\midrule
\multirow{2}{*}{0.1}& fM  & 55.31 & 49.83 & 34.02 & \textbf{64.83} & 16.57 & 66.41 & 52.24 & \textbf{69.52}  \\ 
                    & C10 & 10.06 & \textbf{22.55} & 14.50 & 20.49  & 10.00 & 15.33 & 10.00 & \textbf{29.81} \\ \midrule
\multirow{2}{*}{0.3}& fM  & 64.22 & 57.52 & 10.00 & \textbf{73.55} & 16.26 & 58.07 & 10.00 & \textbf{61.12} \\ 
                    & C10 & 24.24 & 12.59 & 10.00 & \textbf{27.81} & 10.98 & 22.61 & 10.00 & \textbf{38.27}\\ \midrule
\multirow{2}{*}{0.5}& fM & 62.37 & 58.20 & 10.00 & \textbf{76.41}  & 18.36 & 62.49 & 10.00 & \textbf{69.05}\\ 
                    & C10 & 23.99 & 17.28 & 10.06 & \textbf{34.32} &  9.87 & 27.83 & 10.00 & \textbf{37.27}\\ 
\bottomrule
\end{tabular}
\label{tab:untargeted_diff_fraction}
\end{center}
\vspace{-5mm}
\end{table}

\section{Limitation} \label{sec:limitation}

While FedCC demonstrates strong performance against various poisoning attacks in both IID and non-IID settings, several limitations remain.

First, our experiments focus on lightweight CNN architectures (e.g., LeNet and custom CNNs) and relatively small-scale vision datasets such as fMNIST, CIFAR10, and CIFAR100. While these choices reflect practical FL deployments on resource-constrained devices, the generalizability of FedCC to larger-scale datasets (e.g., ImageNet) or more complex architectures (e.g., ResNet, ViTs) remains untested. Additionally, FedCC currently assumes homogeneous model architectures across clients. Extending it to heterogeneous or multi-task client settings is an important direction for future work.

Second, while FedCC does not rely on explicit access to raw data and has shown strong empirical robustness, including against adaptive attacks like DBA, it lacks formal guarantees under broader adversarial conditions. Theoretical analysis of its robustness bounds, especially under targeted mimicry of benign PLR distributions, remains open. Our theoretical insight provides a preliminary foundation but invites further exploration.

Third, while Kernel CKA offers strong representational comparison performance, it incurs computational overhead during aggregation. Although this cost is manageable in our experimental setting, optimizing its computation or exploring more efficient alternatives would be necessary for scalability in large-scale deployments.

\section{Conclusion and Future Work} \label{Section7}

FL has emerged in response to growing concerns about data privacy in collaborative machine learning. However, the distributed nature of FL introduces vulnerabilities to model poisoning attacks, especially under non-IID data settings. While many existing defenses are effective under specific threat models, they often neglect the challenges introduced by client heterogeneity.

In this paper, we proposed FedCC, a robust aggregation method that leverages Kernel CKA to measure representational similarity in the penultimate layer of client models. Through extensive experiments across multiple datasets and attack types, we demonstrated that FedCC effectively mitigates both untargeted and targeted (backdoor) attacks, even under severe non-IID conditions.

For future work, we plan to provide theoretical guarantees for FedCC’s robustness and extend it to scenarios with heterogeneous model architectures. We also aim to evaluate its generalizability on larger and more complex datasets (e.g., ImageNet) and explore its effectiveness against stealthier or adaptive attack strategies. Lastly, we will investigate optimization techniques to reduce the computational cost of Kernel CKA, enabling scalable deployment in large-scale federated systems.

% \section{Conclusion and Future Work} \label{Section7}
% FL has emerged in response to the rising concerns about privacy breaches while employing AI techniques. FL, however, is vulnerable to model poisoning attacks due to a server's blindness to local datasets, making it difficult to assume data distribution and model parameter integrity. In line with this, many robust aggregation algorithms and defenses have been proposed; still, existing approaches are orthogonal to various attacks and lack consideration of non-IIDness. Throughout the exhaustive experiments, FedCC mitigates both untargeted and targeted (or backdoor) attacks while demonstrating its effectiveness in non-IID data environments. We leave theoretical guarantees of FedCC and experiments when clients have different neural network architectures or tasks as future work. 

% \section*{Acknowledgements}
% This research was supported by Healthcare AI Convergence Research \& Development Program through the National IT Industry Promotion Agency of Korea (NIPA) funded by Ministry of Science and ICT (No. S1601-20-1041).

\bibliographystyle{splncs04}
\bibliography{main}

\end{document}